 \documentclass[preprint]{aastex}

\newcommand{\be}{\begin{equation}}
\newcommand{\ee}{\end{equation}}

\newcommand{\Te}   {\mbox{$T_{e}$}}
\newcommand{\zbar} {\mbox{$\bar{z}$}}
\newcommand{\DzT} {\mbox{$D(\bar{z},T_{e})$}}

\slugcomment{ApJLett, in press}

\shorttitle{Approximation to Trace Element Cooling}
\shortauthors{Benjamin, Benson, Cox}

\begin{document}

\newlength{\bigpicsize}
\setlength{\bigpicsize}{4.5in}
\newlength{\smpicsize}
\setlength{\smpicsize}{3.5in}

%% LaTeX will automatically break titles if they run longer than
%% one line. However, you may use \\ to force a line break if
%% you desire.

\title{A Useful Approximation to the Cooling Coefficient of Trace Elements}

%% Use \author, \affil, and the \and command to format
%% author and affiliation information.
%% Note that \email has replaced the old \authoremail command
%% from AASTeX v4.0. You can use \email to mark an email address
%% anywhere in the paper, not just in the front matter.
%% As in the title, you can use \\ to force line breaks.

\author{Robert A. Benjamin, Bradford A. Benson\altaffilmark{1}, Donald P.
Cox}
\affil{Department of Physics, University of Wisconsin-Madison, 
1150 University Ave., Madison, WI 53706}

%\altaffiltext{1} E-mail address: \email{benjamin@physics.wisc.edu}
\altaffiltext{1}{Current address: Department of Physics, Stanford
University, Stanford, CA 94305}

\begin{abstract}

Radiative cooling is an important ingredient in hydrodynamical models
involving evolution of high temperature plasmas.  Unfortunately,
calculating an accurate cooling coefficient generally requires the
solution of over a hundred differential equations to follow the
ionization. We discuss here a simple 2-parameter
approximation for the cooling coefficient due to elements heavier than
H and He, for the temperature range $T= 10^{4}-10^{8}$K.  Tests of the
method show that it successfully tracks the ionization level in severe
dynamical environments, and accurately approximates the
non-equilibrium cooling coefficient of the trace elements, usually to
within 10\% in all cases for which cooling is actually important.  The error
is large only when the temperature is dropping so rapidly due to
expansion that radiative cooling is negligible, but even in this
situation, the ionization level is followed sufficiently accurately.
The current approximation is fully implemented in publicly available
FORTRAN code.  A second paper will discuss general approaches to
approximation methods of this type, other realizations which could be
even more accurate, and the potential for extension to calculations of
non-equilibrium spectra.

\end{abstract}

%% Keywords should appear after the \end{abstract} command. The uncommented
%% example has been keyed in ApJ style. See the instructions to authors
%% for the journal to which you are submitting your paper to determine
%% what keyword punctuation is appropriate.

\keywords{atomic data --- radiation mechanisms:thermal --- plasmas --- hydrodynamics --- supernova remnants --- ISM:general }

%% From the front matter, we move on to the body of the paper.
%% In the first two sections, notice the use of the natbib \citep
%% and \citet commands to identify citations.  The citations are
%% tied to the reference list via symbolic KEYs. The KEY corresponds
%% to the KEY in the \bibitem in the reference list below. We have
%% chosen the first three characters of the first author's name plus
%% the last two numeral of the year of publication as our KEY for
%% each reference.

\section{Introduction}

In this paper we present a reliable but compact
approximation to the high temperature cooling coefficient of heavy
elements in diffuse plasmas under non-equilibrium conditions.\footnote{A set of FORTRAN routines to implement this approximation are provided at ftp://wisp5.physics.wisc.edu/pub/benjamin/BBC. Documentation is given in the file README.bbc.} 
  Such an
approximation is essential for use with multi-dimensional hydrodynamics codes
where the additional burden of following the details of ionization evolution severely restricts the available
spatial resolution.

With only a few exceptions, large hydrodynamic models that incorporate
radiative cooling characterize the cooling coefficient with a single
parameter, the temperature, $L(T_e)$, where the total emissivity per
unit volume is $\Lambda(T_e)=n_{e}n_{H}L(T_{e})~{\rm (ergs~
s^{-1}~cm^{-3})}$ and $n_{e}$ and $n_{H}$ are the electron and
hydrogen densities.  These cooling functions are determined by
assuming either that the ionization state at a given temperature is
characterized by collisional equilibrium, or that all gas follows a
particular pre-calculated ionization history (Shapiro \& Moore 1976;
Edgar \& Chevalier 1983; Sutherland \& Dopita 1993).

Because the cooling of a plasma depends on the ionization history of
the constituent ions, there can actually be a large range in the value
of $L$\ at a given $T_e$, depending upon the details of the ionization
evolution.  We demonstrate, however, that cooling due to trace
elements (those heavier than helium) can be 
approximated by following the evolution of just a single additional
parameter,  the mean charge on the trace ions,
$\bar{z}=\left( \Sigma_{Z,z}A_{Z}~z~c_{Z,z}\right)/\Sigma_{Z} A_{Z}$.
Here, Z is the element number, z is the ionic charge, $A_{Z}$ is the
(linear) abundance of element Z relative to hydrogen, and
$c_{Z,z}=n_{Z,z}/n_{Z}$ is the concentration of a given ion (the
fraction of element $Z$\ with charge $z$).  Because the abundance is
dominated by oxygen, $\bar{z}$\ ranges from zero to about
nine. 

Our method generalizes from a cooling curve, $L(T_e)$, to a cooling
plane $L(T_e,\bar{z})$.  It consequently requires 
another function that allows us to update the ionization level,
$D(T_{e},\bar{z})=(1/n_{e})(d\bar{z}/dt)$.  With the mean
charge as our ionization level indicator,  $D(T_{e},\bar{z})$\ is the difference between the
mean ionization and recombination functions.  Thus
$D(T_{e},\bar{z})=I(T_{e},\bar{z})-R(T_{e},\bar{z})$, where
$I(T_{e},\bar{z})=\left(
\Sigma_{Z,z}A_{Z}~i_{Z,z}(T_{e})~c_{Z,z}\right)/\Sigma_{Z} A_{Z}$\ and
$R(T_{e},\bar{z})=\left(
\Sigma_{Z,z}A_{Z}~r_{Z,z}(T_{e})~c_{Z,z}\right)/\Sigma_{Z} A_{Z}$, and
where $i_{Z,z}(T_{e})$\ and $r_{Z,z}(T_{e})$\ are the ionization and
recombination rate coefficients, respectively, at $T_{e}$\ for stage
$z$\ of element $Z$.  The corresponding cooling coefficient is
$L(T_{e},\bar{z})= \Sigma_{Z,z}A_{Z}~j_{Z,z}(T_{e})~c_{Z,z}$, where
$j_{Z,z}(T_{e})$\ is the cooling coefficient per ion $Z, z$.

To implement this approximation we search for a
reasonable description of the nonequilibrium ionization
concentrations, $c_{Z,z}(T_{e},\bar{z})$, 
representative of those found at $T_{e}$\ and $\bar{z}$.  This
assumes that under actual nonequilibrium conditions, the
distribution over ion states will depend less on the details of
the past history than on how far the
present mean charge differs from the collisional equilibrium value at
the current temperature.  In this paper, 
we consider the
ionization distributions that arise in isothermal relaxation to
equilibrium for each temperature $T_e$, starting with  
nearly fully ionized or fully neutral gas.

The atomic data used to develop the manifolds comes from Raymond \&
Smith (1977), with updates described in Raymond \& Cox (1985),
corrections in oscillator strengths in the cooling transitions of
Li-like ions (Shull \& Slavin 1994), and revised dielectronic
recombination rates of Romanik (1988). The abundances are taken from
Anders \& Grevesse (1989).  The effects of charge exchange are not
included; errors introduced by neglecting this 
are smaller than the uncertainties in the atomic
data and elemental abundances used to generate the cooling curve.

\label{secintro}

\section{A Manifold of Isothermal Evolutions}

We have performed the manifold of non-equilibrium isothermal
evolutions described below to form the basis for our approximation,
producing tables of the cooling coefficient of trace elements
$L(T_e,\bar{z})$ and our ionization evolution functions
$I(T_e,\bar{z})$\ and $R(T_e,\bar{z})$.

Gas is initialized with equilibrium ionization appropriate
to some initial temperature $T_o$.  Its temperature is then suddenly
changed to $T_e$ and held fixed as the ionization evolves to equilibrium.  By choosing $T_o$ to be low (e.g., $10^{4.0}$
K), and performing the calculation for a dense set of temperatures
$T_e$ between $10^{4.0}$ and $10^{8.0}$ K, the runs sample the full
range of conditions possible in under-ionized gases.  By repeating the
whole set once again with very high $T_o$ (e.g. $10^{8.0}$ K),
conditions representing over-ionized gases are explored.  Together
these cases sample the full range of $\bar{z}$. 

A particular moment in an evolution can
be characterized by $T_o$, $T_e$, and the fluence ($f=\int n_e dt$).
The ionization state is known, allowing
straightforward evaluation of the mean charge $\bar{z}$ and then 
using \zbar~ as the index of the time evolution.
Evolution at constant temperature makes it easy to acquire
the tables of $L(T_e,\bar{z})$, $I(T_e,\bar{z})$, $R(T_e,\bar{z})$, and $D(T_e,\bar{z})$  on a fixed grid of $T_e$ and \zbar. 

\begin{figure}[ht!]
\includegraphics[angle=90,totalheight=\bigpicsize]{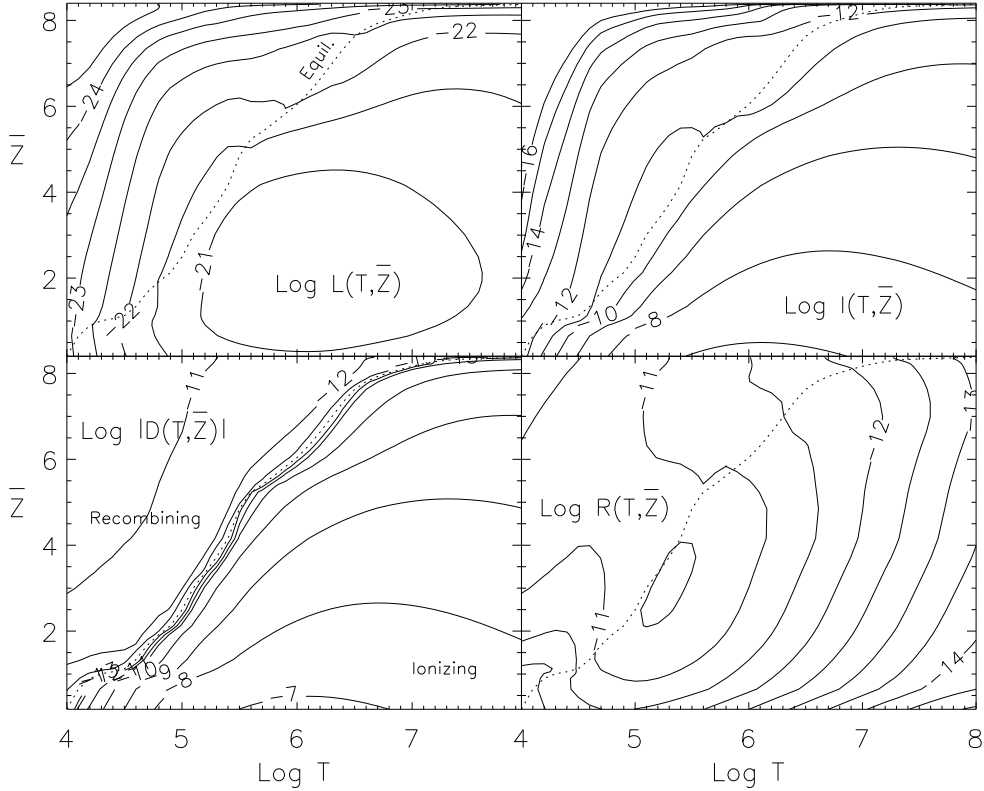}
\caption{ Contours of the trace element cooling coefficient,
$L(T_{e},\bar{z})~{\rm (erg ~cm^{3}~s^{-1})}$ [top left panel], the
mean ionization rate coefficient, $I(T_{e},\bar{z}){\rm
~(cm^{3}~s^{-1})}$ [top right panel], the mean recombination rate
coefficient, $R(T_{e},\bar{z}){\rm ~(cm^{3}~s^{-1})}$ [bottom right 
panel], and the absolute value of the ionization evolution function,
$|D(\zbar,\Te)| {\rm~( cm^{3}~s^{-1})}$ [bottom left panel], as a
function of electron temperature and mean charge $\zbar$. The dotted
line in each panel shows the mean charge in collisional equilibrium as
a function of gas temperature, $z_{eq}(T_{e})$. The function
$|D(\zbar,\Te)| {\rm~( cm^{3}~s^{-1})}$ passes through zero along this
line. Above the equlibrium curve, the gas is recombining and
D(\zbar,\Te) is negative; below, it is ionizing and D(\zbar,\Te) is
positive. These functions were calculated for a grid of isothermal
evolutions.  Gas was either started with ionization fractions
corresponding to $T_{o}=10^{4}$ K or $T_{o}=10^{8}$ K. It then ionized
up or recombined down to $\bar{z}_{eq}(T_{e})$ at fixed temperature
$T_e$. The resultant rate coefficients were calculated versus
$\bar{z}$\ for each $T_e$ for the range $T_e = 10^{4}-10^{8}$K.  }
\label{fig1}
\end{figure}

The cooling function shown in Figure 1 shows a huge peak at $T_{e} \sim 10^{6}$
K and $\bar{z} \sim 2$ due to collisional excitation of low stages of
ionization. It falls rapidly to the left due to Boltzmann factors, and gradually to the right due to decreasing excitation cross-sections. It falls at higher \zbar~ as
fewer bound electrons with low enough energy are available for
collisional excitation.  Above the equilibrium line the gas is
over-ionized and collisional excitation is difficult, particularly at
lower temperatures. At a given $\bar{z}$, the mean ionization function, I, \
behaves much like the collisionally excited cooling, with vagaries at
 the transition from $\zbar=1$ to 2, and at $\bar{z} \cong 6$ where the mean ionization rate is
sensitive to the relative proportions of helium- and lithium-like
oxygen.  

The structure of the mean recombination rate, R,  is
more gradual.  In the upper left corner it is dominated by radiative
recombination and the gradual increase with $\bar{z}$\ and decrease
with temperature are as expected.  Deviations from that smooth pattern
in the rest of the diagram are due to dielectronic recombination.  In
both I and R, there are distortions in the patterns just
above the equilibrium line. These arise because, in relaxing to equilibrium,
a recombining plasma goes through a considerable compaction of its
ionization distribution.

The principal feature of the rate function
\DzT~ is that ionization toward equilibrium from
below is much faster than recombination toward it from above. With
increasing distance from equilibrium, the ionization rate increases
dramatically, the recombination rate only gradually.

\section{Testing With Cooling Blast Wave Evolutions}

To test the accuracy of our approximation for a wide range of
situations, a representative set of cases were examined in which
single parcels of gas in collisional equilibrium at $10^4$\ K were
suddenly shock heated and then subjected to varying degrees of
expansion.  The test situation was modeled on the Sedov blast wave
solution for an explosion of energy $10^{51}$
ergs into a homogeneous medium of particle density $n_o$.  By varying the
preshock density and the assumed distance of the parcel from the
explosion site, we adjusted the post shock temperature and the
timescales for depressurization versus radiative cooling. These
scenarios test the two behaviors in which drastic
departures from equilibrium occur: the rapid ionization of gas passing
through a shock front and the rapid adiabatic cooling of an
over-ionized plasma.
  
Each parcel's evolution can be fully characterized by $n_o$ and
$T_2$. We have approximated $p(t)=p_2
(t_2/t)^\alpha$, with $t_2$ being the time the parcel is shocked, and $\alpha=1.9$ determining the rate of
decompression.  The temperature evolution is then

 \begin{equation} 
\frac{dT}{dt} = -\frac{2}{5}T\bigg(\frac{L(T, \zbar)n_e n_H}{p} + \frac{{\alpha}}{t}\bigg). 
\label{Trate}
\end{equation} 
where  $n_H$ is the particle density of hydrogen (ionized and neutral).

We have carried out each simulation twice,  first, solving the full ionization balance evolution exactly, then using our cooling approximation.  In the latter, we solve the
ionization evolution for hydrogen, helium, and \zbar~ (four rate
equations). The cooling is the sum of that from hydrogen, helium, and
our tabulated cooling function. 

In order to test the widest possible range of conditions, we chose
three values of $T_2$, with three different values of $n_o$ for
each. The values of density were spaced to alter the initial ratio of
depressurization to radiative cooling from approximately 0.4 at the
high density end, to 6 and then 4000 at the lower densities.  For the
densest runs, the structure is very similar to that of a steady state
radiative shock.  For the next lower density cases, depressurization
has strong effects but cooling is still sufficiently rapid that the
density eventually rises rather than falling with time.  At the lowest
densities, the density and temperature fall adiabatically and the
ionization level is soon frozen in.

\begin{figure}[ht!]
\includegraphics[angle=90,totalheight=\bigpicsize]{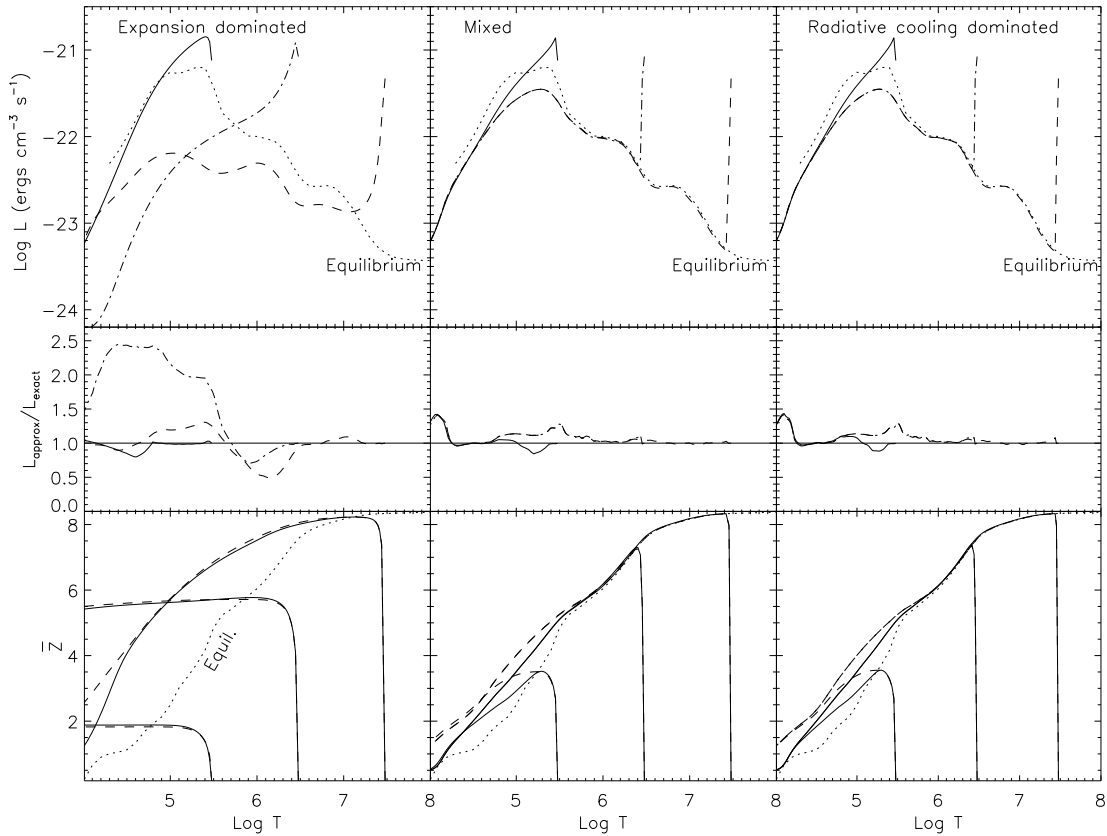}
\caption{Comparison of cooling and ionization evolution for nine
cases using the full ionization evolution and the approximation
presented in this paper. The top row shows the evolution of the exact
trace element cooling coefficient versus temperature. The left panel
shows expansion-dominated evolution with 
$(n_{o},T_{2})= (1~{\rm cm^{-3}},3 \times 10^{7}~{\rm K})$ (dashed),
$(3\times 10^{-4}~{\rm cm^{-3}},3 \times 10^{6}~{\rm K})$ (dash-dot), and 
$(10^{-7}~{\rm cm^{-3}},3 \times 10^{5}~{\rm K})$ (solid); 
the central panel shows ``mixed'' evolution with 
$(n_{o},T_{2})= (2 \times 10^{4}~{\rm cm^{-3}},3 \times 10^{7}~{\rm K})$ (dashed),
$(10~{\rm cm^{-3}},3 \times 10^{6}~{\rm K})$ (dash-dot), and 
$(2 \times 10^{-3}~{\rm cm^{-3}},3 \times 10^{5}~{\rm K})$ (solid); 
the right panel shows radiative cooling dominated evolution with
$(n_{o},T_{2})= (10^{6}~{\rm cm^{-3}},3 \times 10^{7}~{\rm K})$ (dashed),
$(300~{\rm cm^{-3}},3 \times 10^{6}~{\rm K})$ (dash-dot), and 
$(0.1~{\rm cm^{-3}},3 \times 10^{5}~{\rm K})$ (solid).
The time
evolution for each curve is from right to left. The
cooling curve for collisional ionization equilibrium is also shown (dotted). The
middle set of panels shows the ratio of the approximation for trace
element cooling to the exact value for each of the above cases. For
``mixed'' and ``radiative'' evolution, the highest temperature cases
are sufficiently similar that the curves overlie each other. The
lowest set of panels show the ionization evolution for these cases in
the T-\zbar~ plane. The solid curves show the exact evolution for the
nine cases above, while the dashed lines show the slight departures
that result by using the approximation we present. The dotted line in
each panel shows the mean charge in collisional equilibrium as a
function of gas temperature, $z_{eq}(T_{e})$, as in Figure 1.}
\label{fig2}
\end{figure}

 Figure 2 shows the exact trace element cooling coefficient evolutions
versus temperature, the ratio of our approximate results to the exact
ones, and the evolutions of mean charge, both exact and approximate.
In each case, the cooling function begins very high just after the
shock, when the gas is still briefly nearly neutral.  At the two
higher temperatures in the middle and right hand panels, the gas
rapidly ionizes up and the cooling function drops precipitously to
very close to the equilibrium value.  This ``ion flash" is so
rapid that the integrated cooling during this period is negligible, as evidenced by the nearly vertical tracks.  Thereafter,
these tracks follow the equilibrium curve down to just below $10^6$K,
after which their recombination cannot keep up with the cooling and
the cooling rate falls below the equilibrium value.  In the lowest temperature cases, significant
cooling takes place before the cooling function drops through
equilibrium, after which it plunges toward the nonequilibrium curve of
the initially higher temperature cases.  The differences between the
radiation dominated and mixed evolutions are almost negligible.  In
the expansion dominated cases, however, things are quite
different.   The temperature 
drops so rapidly that the gas is soon over-ionized; recombination is slower
than the temperature drop,  and the ionization
level declines only gradually or is frozen in.

The middle row of panels shows the ratio of approximate to exact cooling 
coefficients.  In the two right hand panels, the approximation is good to 
within about 10\% except below $10^{4.2}$K where it is consistently high 
by about 40\%, and in a peak at $10^{5.5}$K where it can be as much as 20 
to 30\% high. At this  temperature, the 
approximate $\bar{z}$\ evolutions of the two higher temperature cases in the two 
right lower panels of Figure 2 diverge by $+0.4$ charge states from the exact results. This is due in part to the excess cooling, but the 
recombination rate is almost certainly slightly too low as well. 

For the rapidly depressurizing cases, radiative cooling is inconsequential. Nevertheless, the cooling coefficient which differs substantially from the other cases, is still reasonably well fit, except for the
intermediate temperature case for which $\bar{z}$\ gets frozen at a
value of 5.5.  In that case, the value of $\bar{z}$\ is well
represented by the approximation, but the approximate cooling
coefficient is too high.  The freezing in of the ionization
is extremely well matched, except very late in the highest initial temperature case.   
 
We have also compared the time evolution of density, temperature, mean
charge, and the cooling coefficients for the exact calculation, our 
approximation, and the case where the trace elements are assumed to 
be in collisional ionization equilibrium. While the time 
evolution of temperature and density of the exact calculations and
our approximation are in reasonably good agreement, the case with
collisional ionization equilibrium shows significant differences in 
the time evolution of temperature. In addition, the ionization 
equilibrium assumption provides a poor estimate of the true ionization level
of the gas, especially for the expansion dominated cases.

\section{Summary and Conclusions}

We  provide a means by which
large hydrocodes can include an accurate approximation to the
radiative cooling coefficient, one far more responsive to the vagaries
of dynamical environments than any single function of temperature. 
The method also
follows the mean ionization level of the gas.  Both the trace element cooling coefficient $L(T_e,\bar{z})$, and the
rate of change of $\bar{z}$\ depend only on the identification of a
manifold of representative ionization concentrations found at a particular temperature and mean charge.  The 
ionization concentrations we used came from the
isothermal relaxations of both highly over-ionized and highly
under-ionized gas. 
In the test cases presented, both the cooling coefficient and mean charge evolution are
quite accurately approximated.  For all cases in which radiative
cooling is significant, the approximation never errs by more than
30\%.  The error is usually less than 10 to 15\%, well within the
accuracy of the true cooling coefficient, which is limited by
uncertainties in atomic data and abundances.  The error in the cooling coefficient
can be somewhat larger in examples with extreme amounts of
depressurization, but never by enough to make the negligible radiative
cooling appear to be significant.

Our approximate cooling coefficients and charge evolution rates,
however, showed patterns in their modest errors, patterns which we
believe we understand and can eliminate with future work.  
Should a potential user wait for these improvements before beginning
to implement this method?  The current model provides an excellent
approximation for the cooling coefficient. If one is interested only
in dynamics, it is certainly sufficient. Implementation of the next
generation requires only swapping one set of tables for another. Our
future work will also examine the possibility that the ion
distributions can be used to provide absorption and emission spectra
in post processing. It is possible to do this at the two parameter
level, but higher order corrections will also be examined. The latter
are expected to be two to four times more complex, and will be pursued
only if spectral accuracy requires it.

An additional consideration for gas with $T \sim 10^{4}~K$ and 
$\bar{z} ^{<}_{\sim} 2$ is the effect of photoionization on the ionization 
balance and therefore the cooling. Our scheme can be easily modified to 
incorporate photoionization of hydrogen and helium. However, the effects of 
photoionization of trace elements would require a table of photoionization 
correction factors for $L(T_{e},\bar{z})$ and $D(T_{e},\bar{z})$. This too 
will be considered in future refinements.  

\acknowledgements

We would like to thank John Raymond for compiling and providing much
of the atomic data that went into this work, and for a very careful
reading of the manuscript.  We would also like to thank NASA
Astrophysical Theory grant NAG5-8417 for financial support of this
work. And finally, we would like to acknowledge the valuable
contributions of Leo Krzewina, Angela Klohs, Andrew Pawl, and Tim
Freyer who all expended some effort on cracking this problem.

\clearpage

%% No more than seven \figcaption commands are allowed per page,
%% so if you have more than seven captions, insert a \clearpage
%% after every seventh one.

%% There must be a \figcaption command for each legend. Key the text of the
%% legend and the optional \label in curly braces. If you wish, you may
%% include the name of the corresponding figure file in square brackets.
%% The label is for identification purposes only. It will not insert the
%% figures themselves into the document.
%% If you want to include your art in the paper, use \plotone.
%% Refer to the on-line documentation for details.

%% Tables should be submitted one per page, so put a \clearpage before
%% each one.

%% Two options are available to the author for producing tables:  the
%% deluxetable environment provided by the AASTeX package or the LaTeX
%% table environment.  Use of deluxetable is preferred.
%%

%% Three table samples follow, two marked up in the deluxetable environment,
%% one marked up as a LaTeX table.

%% In this first example, note that the \tabletypesize{}
%% command has been used to reduce the font size of the table.
%% Note also that the \label command needs to be placed 
%% inside the \tablecaption.

%% The following command ends your manuscript. LaTeX will ignore any text
%% that appears after it.

\end{document}